\documentclass[english]{emulateapj}
\usepackage[T1]{fontenc}
\usepackage[latin1]{inputenc}
\makeatletter
\usepackage{graphicx}
\usepackage{amssymb}
\usepackage{times}
\usepackage{babel} 

\newcommand{\peryr}{\,{\rm yr^{-1}}}
\newcommand{\pers}{\,{\rm s^{-1}}}
\newcommand{\erg}{\,{\rm erg}}

\newcommand{\km}{{\,\rm km}}
\newcommand{\Gyr}{{\,\rm Gyr}}
\newcommand{\Myr}{{\,\rm Myr}}
\newcommand{\pc}{\,\mathrm{pc}}

\newcommand{\Mo}{M_{\odot}}

\bibpunct{\hspace{-2pt}}{}{}{a}{}{} 

\shorttitle{RED-SHIFTS OF SHORT GRBs}
\shortauthors{HOPMAN ET AL.}

\makeatother
\begin{document}

\title{The redshift distribution of short $\gamma$-ray bursts from dynamically formed \\
neutron star binaries}

\author{Clovis Hopman\altaffilmark{1}, Dafne Guetta\altaffilmark{2,3},
Eli Waxman\altaffilmark{1} and Simon Portegies
Zwart\altaffilmark{4,5}}

\altaffiltext{1}{Faculty of Physics, Weizmann Institute of
Science, POB 26, Rehovot 76100, Israel}
\altaffiltext{2}{Osservatorio astronomico di Roma, v. Frascati 33,
00040 Monte Porzio Catone, Italy}
\altaffiltext{3}{Einstein Minerva
center for theoretical physics, Weizmann Inst.}
\altaffiltext{4}{Astronomical Institute 'Anton Pannekoek', University
of Amsterdam, Kruislaan 403, Netherlands} 
\altaffiltext{5}{Section
Computational Science, University of Amsterdam, Kruislaan 403,
Netherlands}
\email{clovis.hopman@weizmann.ac.il}

\begin{abstract}
Short-hard $\gamma$-ray bursts (SHBs) may arise from gravitational
wave (GW) driven mergers of double neutron star (DNS) systems. DNSs
may be "primordial" or can form dynamically by binary exchange
interactions in globular clusters during core-collapse.  For
primordial binaries, the time delay between formation and merger is
expected to be short, $\tau\sim0.1$~Gyr, implying that the redshift
distribution of merger events should follow that of star-formation.
We point out here that for dynamically formed DNSs, the time delay
between star-formation and merger is dominated by the cluster
core-collapse time, rather than by the GW inspiral time, yielding
delays comparable to the Hubble time.  We derive the redshift
distribution of merger events of dynamically formed DNSs, and find it
to differ significantly from that typically expected for primordial
binaries. The observed redshift distribution of SHBs favors dynamical
formation, although a primordial origin cannot be ruled out due to
possible detection biases. Future red-shift observations of SHBs may
allow to determine whether they are dominated by primordial or
dynamically formed DNSs.
\end{abstract}


\section{Introduction}

Observations of $\gamma$-ray bursts (GRBs) indicate that they divide
into two classes (Kouveliotou et al.  \cite{Kea93}). GRBs of one class
are of relatively long ($\gtrsim2\!-\!200$ sec) duration and have
softer spectra. Long-soft GRBs occur in star forming galaxies with
high redshift $z$ (van Paradijs et al. \cite{Par97}), and their
association in several cases with type Ibc SNe (\cite{G98, Stanek03,
Hjorth03, Malesani04, Campana06}) suggests that they are the result of
core collapse SN explosions of massive stars
(\cite{Woosley93,Pac98,MacFadyen99}). The second class of GRBs have
short duration ($<2$ sec) and harder spectra. Afterglows of short-hard
GRBs (SHBs) have only recently been observed (e.g., Gehrels et al.;
\cite{Gea05}; Bloom et al. \cite{Bea05}; Berger et al. \cite{BPea05};
Fox et al. \cite{Gea05}), and this has led to the first
identifications of SHB host galaxies. In contrast to long GRBs, SHBs
were found to occur in at least some cases in elliptical galaxies with
very low star formation rates (SFRs), of order
$\lesssim0.1\Mo\peryr$. It is therefore unlikely that the progenitors
of SHBs are also massive stars, since these have very short
($\sim\,{\rm few}\,\Myr$) life times.

The gravitational wave (GW) driven merger of a double neutron star
(DNS) may lead to a SHB (Goodman 1986; Paczy\'nski 1986; Eichler et
al. \cite{ELP89}; Narayan, Paczynski \& Piran \cite{NPP92}). If the
time-lag $\tau$ between DNS formation and the merger is large
(comparable to a Hubble time), it forms a natural explanation why SHBs
occur in galaxies where the SFR is very low.

We consider two formation mechanisms for DNSs. If two massive stars
are born as a binary system, a DNS may form after the super nova (SN)
explosion of both components (``primordial'' DNSs). Population
synthesis models of massive binaries yield typically short merger
times ($\tau\!\approx\!0.1\Gyr$), implying that the SHB
$z$-distribution function (DF) closely follows the SFR history (e.g.,
Bloom, Sigurdsson \& Pols \cite{BSP99}; Belczynski et
al. \cite{Bel01}, \cite{Bel06}). The kick-velocity of the DNS is low
($\lesssim50{\rm km\,s^{-1}}$; Dewi, Podsiadlowski \& Pols
\cite{DPP05}), and the spatial distribution of SHB mergers should
follow the light distribution of their host galaxy (Portegies Zwart \&
Yungelson \cite{PY98}; Bloom et al. \cite{BSP99}).

Another possibility is that at the moment of star formation the
neutron stars (NSs) are {\it not} in the same binary system, but one
of the NSs is in a binary with a low mass main sequence (MS) star. In
globular clusters (GCs), such binaries are likely to have an exchange
interaction with a single neutron star (Sigurdsson \& Phinney
\cite{Sig95}; Efremov \cite{Efr00}; Grindlay, Portegies Zwart \&
McMillan \cite{GPZM06}, hereafter GPZM06), and thus form DNSs. A
significant fraction ($\sim30\%$) of all NS mergers in the Universe
may stem from such dynamically formed systems (GPZM06).

For exchange interactions to occur, the stellar density must be very
high. The delay time for dynamically formed DNSs is therefore mainly
determined by the time until core-collapse (CC) of GCs, which is
typically comparable to the Hubble time (\S\ref{s:tau}). We show that
the predicted $z$-DF of SHBs is different for primordial and
dynamically formed DNSs, so that future $z$ observations may determine
which formation channel (if any) is dominant. To date, the few SHBs
with $z$-detection favor dynamical formation, although possible biases
and small number statistics make a final conclusion at this point
impossible (\S\ref{s:Nz}). We discuss our results in
\S\ref{s:discussion}.

\section{Distribution of merger times}\label{s:tau}

The delay time $\tau$ is a sum of the time $t_{cc}$ until the
dynamical formation of a DNS during core-collapse, and the time
$t_{\rm GW}$ until the DNS merges. Here we determine the resulting
delay function $(dp/d\tau)_{\rm dyn}$ of dynamically formed DNSs.

GPZM06 performed scattering experiments with {\tt scatter3} and {\tt
sigma3} in the {\tt Starlab} environment (Portegies Zwart et
al. \cite{Por01}) to determine the cross-section for the formation of
NS binaries when a NS interacts with a (NS, MS) binary. We use the
orbital parameters of the resulting binaries to determine the
resulting DF of GW merger times $dp_{\rm GW}/dt_{\rm GW}$, which was
found to be well fitted by $dp_{\rm GW}/dt_{\rm GW}\propto t_{\rm
GW}^{-1.1}$. Typical merger times $t_{\rm GW}$ are very short compared
to the Hubble time.

The DNS formation rate per GC can be estimated as $\Gamma\!\approx\!
4\,{\rm Gyr^{-1}}\,n_6 v_1(N_{\rm pr}/20)$, where
$n\!=\!10^6\pc^{-3}n_{6}$ is the number density of NS stars,
$v\!=\!10\km\pers v_1$ is the velocity dispersion, and $N_{\rm pr}$ is
the number of progenitor binaries containing one NS (GPZM06). Since
$\Gamma\!\propto\! n$, DNSs form when the GC is very dense, i.e.,
during the CC phase. The delay between the formation of the GC and the
SHB is the sum of the delay time $t_{\rm GW}$ and the time $t_{cc}$
between formation of the GC and CC.

For the DF of CC times we make the following assumptions: the CC times
of the GCs in our Galaxy are representative for the whole Universe;
the relation between $t_{cc}$ and the half mass relaxation time is
only a function of concentration, and it is well approximated by the
relation given by Quinlan (\cite{Q96}); the formation rate of GCs is
proportional to the total SFR. We (conservatively) neglect repeated
phases of CC (``gravothermal oscillations''; Sugimoto \& Bettwieser
\cite{SB83}; Makino \cite{M96}), which would lead to even lower
redshifts.

\begin{figure}
\includegraphics[width=1.0\columnwidth, keepaspectratio]{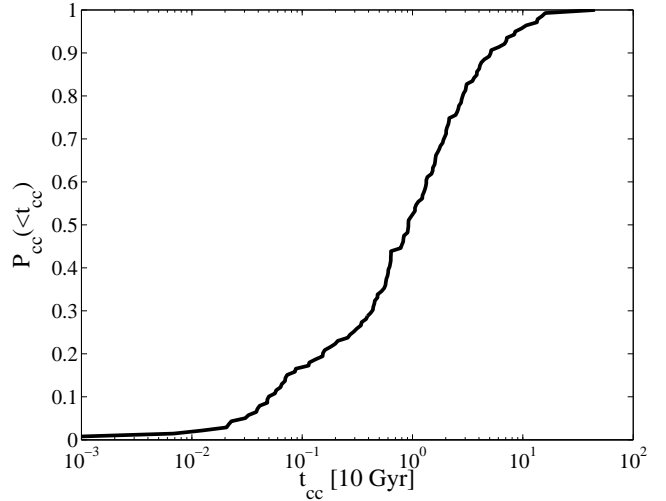}
\figcaption[FileName]{Cumulative DF $P(<t_{cc})$ of the CC times of the
GCs in our Galaxy. Most GCs have CC times that are larger than 1
Gyr.\label{f:tcc}}
\end{figure}

We use the half-mass relaxation times given by Harris (\cite{H96}) to
find the cumulative DF $P_{cc}(<t_{cc})$ for GCs with CC times smaller
than $t_{cc}$ (Fig. \ref{f:tcc}). The time $t_{cc}$ for a GC
between formation and CC is somewhat uncertain. Results in the
literature for single mass systems without binaries agree
approximately (e.g., Quinlan \cite{Q96}; Joshi, Nave \& Rasio
\cite{Joh01}; Baumgardt \cite{Bau01}) . A spectrum of masses can
significantly decrease $t_{cc}$ (G\"urkan, Freitag, \& Rasio
\cite{GFR04}), while primordial binaries increase $t_{cc}$ (Fregeau et
al. \cite{FGR03}). The estimate by Quinlan (\cite{Q96}) leads to most
GCs having $t_{cc}> t_H$ (Fig. \ref{f:tcc}), consistent with the
observation that most $(\sim80\%)$ GCs in the Galaxy have not yet
experienced CC.

\begin{figure}
\includegraphics[width=1.0\columnwidth,keepaspectratio]{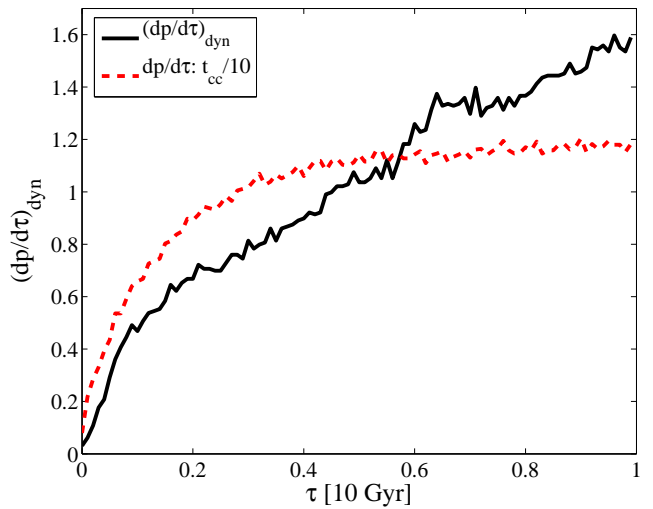}
\figcaption[FileName]{\label{f:delay}The distribution $(dp/d\tau)_{\rm
dyn}$ of inspiral times (black solid line) from the dynamically formed
DNS mergers, combining the delays from CC and GW inspiral. The average
delay time is $\bar{\tau}\approx6\Gyr$. For comparison, we also plot
the DF that would result in the (unlikely) case that $t_{cc}$ would be
10 times smaller than our estimate (red dashed line). Since the delay
function is still a growing function of $t_{cc}$ and has a large
average value ($\bar{\tau}\approx5\Gyr$), the conclusions on the
resulting $z$-DF would not be affected, stressing the robustness of
our results. }
\end{figure}

The probability function of the total time $\tau=t_{\rm GW}+t_{cc}$ to
be in the interval $(\tau, \tau+d\tau)$ is given by
\begin{equation}\label{eq:Ftau}
\left({dp\over d\tau}\right)_{\rm dyn}= {d\over d\tau}\int_0^{\tau}dt_{cc}{dp_{cc}\over dt_{cc}}\int_0^{\tau-t{cc}}dt_{\rm GW} {dp_{\rm GW}\over dt_{\rm GW}}.
\end{equation}
The resulting delay function is shown in figure (\ref{f:delay}); we
weighed the CC times by the GC luminosity. Since the GW inspiral time
DF diverges towards small times, $(dp/d\tau)_{\rm dyn}$ is mostly
determined by the DF of of CC times.  It is also shown in
fig. (\ref{f:delay}) that the delay time DF $(dp/d\tau)_{\rm dyn}$ is
not modified significantly by choosing $t_{cc}$ which is smaller than
that of fig.~\ref{f:tcc} by a factor of 10. This demonstrates the
robustness of the conclusion that for dynamical DNSs, the delay times
are comparable to the Hubble time.

In contrast to the delay function for primordial DNSs, we find that
{\it the delay function of dynamically formed DNSs grows with the
delay time}. For delay times shorter than 10 Gyr, the average delay
time is $\bar{\tau}\approx6\Gyr$.

\section{Redshift distribution}\label{s:Nz}
We derive the predicted $z$-DF of SHBs, which depends on the event
rate of SHBs as a function of $z$, the luminosity DF of SHBs, the
delay function, and the detection threshold.

\subsection{The intrinsic $z$-distribution}

\begin{figure}
\includegraphics[width=1.0\columnwidth, keepaspectratio]{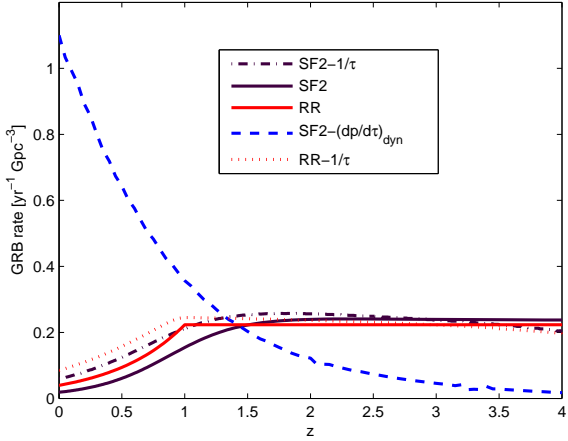}
\figcaption[FileName]{\label{tdelay} Intrinsic SHB rates as a function
of $z$. For primordial binaries, the SHB $z$-DF (dotted and
dash-dotted lines) is only slightly shifted to lower $z$, compared to
the SFR distribution (solid lines), by the delay function of
primordial DNSs.  SHBs from mergers of dynamically formed DNS occur at
much lower red-shift due to the large delay (dashed line).}
\end{figure}

For DNS mergers, the {\it intrinsic} (as opposed to observed) SHB rate
is given by the convolution of the star formation rate ${\rm SFR}(z)$
with the distribution $dp(\tau)/d\tau$ of time delays,

\begin{equation}
\label{rSHB}
N_{\rm intr}(z)\propto \int_z^{\infty}dz'{dt\over dz'}{\rm
SFR}(z'){dp\over dt}\left[t(z)-t(z')\right]
\end{equation}
(e.g. Guetta \& Piran (\cite{GP05} [hereafter GP05]; Nakar, Gal-Yam \&
Fox \cite{NGF06}); $t(z)$ is the age of the Universe as function of
$z$. We employ the SF2 model of Porciani \& Madau (2001)
\begin{eqnarray}
\label{SFR} 
{\rm SFR}_{\rm PM}(z)  \propto { 23 e^{3.4z}[\Omega_M(1+z)^3+\Omega_k(1+z)^2+\Omega_{\Lambda}]^{1/2} \over({e^{3.4z}+22}) (1+z)^{3/2}},\nonumber\\
&&
\end{eqnarray} 
and the Rowan-Robinson (\cite{Row99}) SFR,

\begin{equation}
\label{RR}
{\rm SFR}_{\rm RR}(z) \propto
\left\{ \begin{array}{ll}
10^{0.75 z} & z<1 \nonumber \\
10^{0.75} & z>1.
\end{array}
\right.
\end{equation}
In figure (\ref{tdelay}) we
show the intrinsic SHB $z$-DF $N(z)$ for $(dp/d\tau)_{\rm
prim}\propto1/\tau$ (as possibly appropriate for primordial DNSs,
Bloom et al. \cite{BSP99}; Belczynski et al. \cite{Bel06}) and
$dp/d\tau=(dp/d\tau)_{\rm dyn}$ (Eq. [\ref{eq:Ftau}]).

\subsection{The SHB luminosity function}

For the {\it observed} $z$-DF, the DF of peak luminosities and the
minimal observable flux $P_{\rm lim}$ are required. Here we follow the
procedure outlined in Guetta, Piran \& Waxman (\cite{GPW04}). We
consider a broken power law peak luminosity function (LF) with lower
and upper limits, $1/\Delta_1$ and $\Delta_2$, respectively:
\begin{equation}
\label{Lfun} \Phi_o(L) d\log L =C_0 d\log L \left\{
\begin{array}{ll}
(L/L^*)^{-\alpha}; & L^*/\Delta_1 \!<\! L \!<\! L^* \\ (L/L^*)^{-\beta}; & L^* \!<\!
L \!<\! \Delta_2 L^*
\end{array}
\right.
\end{equation}
where $C_0$ is a normalization constant. This is the
``isotropic-equivalent" LF, i.e. it does not include a correction
factor due to beaming. Following Schmidt (\cite{S01}) we approximate
the effective spectral index in the observed range of 20 or 50keV to
300 keV as $-1.1$ ($N(E)\propto E^{-1.1}$), and we use
$\Delta_{1,2}=(30,100)$ (GP05).  Both values are chosen such that even
if there are bursts less luminous than $L^*/\Delta_1$ or more luminous
than $\Delta_2 L^*$ they will be only very few (less than about 1\%)
of the observed bursts outside the range $(L^{*}/\Delta_1,L^*
\Delta_2)$.  We find $\alpha=0.6$ and $\beta=2$, and then constrain
$L^{*}$ by comparing the predicted peak flux distribution with the one
observed by BATSE (see Guetta et al. [\cite{GPW04}] and GP05 for
details).  The best fit values of $L^*$ are reported in Table
(\ref{t:fit}).

\begin{table}[t]
\caption{Model parameters and goodness of fit}
\begin{tabular}{lllll}
 \hline
Model & $L^*$ & P$_{\rm lim}$ &KS test$^a$& KS test$^a$\\
   & $[10^{51}\erg\pers]$   &$[{\rm ph\, cm^{-2}\, s^{-1}}]$ &(with z=0.7)&(no z=0.7) \\
  \hline
 SF2-$1/\tau$               & $2 $     & 1    & 0.05      & 0.01\\
 RR-$1/\tau$                & $1.5$    & 1    & 0.09      & 0.03 \\
 SF2-$(dp/d\tau)_{\rm dyn}$ & $0.3$    & 1    & 0.7       & 0.8\\
SF2-$1/\tau$                & $2 $     & 2.5  & 0.15      & 0.05\\
 RR-$1/\tau$                & $1.5$    & 2.5  & 0.25      & 0.09 \\
 SF2-$(dp/d\tau)_{\rm dyn}$ & $0.3$    & 2.5  & 0.4        & 0.6\\
 \hline
\multicolumn{5}{l}{$^a$ The redshifts used are $z$=\{0.225, 0.16, 0.257, 0.55\} for SHBs }\tabularnewline
\multicolumn{5}{l}{\,\,\,\,\{050509, 050709, 050724, 051221\}. We consider separately the case}\tabularnewline
\multicolumn{5}{l}{\,\,\,\, including GRB050813, which is tentatively associated with a cluster }\tabularnewline
\multicolumn{5}{l}{\,\,\,\,  at $z=0.7$ (Gladders et al. \cite{Gla05}).}
\label{t:fit}
\end{tabular}
\end{table}

\subsection{The observed $z$ distribution}

\begin{figure}
\includegraphics[width=1.0\columnwidth, keepaspectratio]{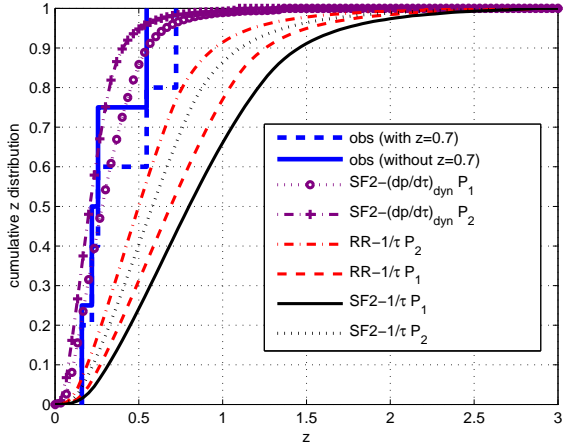}
\figcaption[FileName]{\label{cdf} Observed (histogram) and theoretical
cumulative $z$-DFs for different SFR evolution (SF2 and RR) and
different merger delay times ($1/\tau$ expected for primordial DNSs,
and $(dp/d\tau)_{\rm dyn}$ expected for dynamical DNSs). Model
distributions are shown for peak flux detection threshold $P_{\rm
lim}=P_1\equiv1\,{\rm ph\,cm^{-2}\, s^{-1}}$, which reproduces the
observed detection rate of SHBs, and for $P_{\rm
lim}=P_2\equiv2.5\,{\rm ph\,cm^{-2}\, s^{-1}}$, which reproduces the
observed detection rate of SHBs {\it with $z$ determination} .}
\end{figure}

The expected $z$-DF of the observed bursts is
\begin{equation}
\label{redshift} N_{\rm obs}(z)= \frac{N_{\rm intr}(z)}{1+z} \frac{dV(z)}{dz}
\int_{L_{\rm min}(P_{\rm lim},z)}^{L_{\rm max}} \Phi_o(L)d\log L \
 ,
\end{equation}
where $L_{\rm max}=\Delta_2 L^*=100L^*$ and $L_{\rm min}$ is the
luminosity at $z$ corresponding to the minimum peak flux $P_{\rm lim}$
required for detection. We estimate that $P_{\rm lim}$ for SWIFT is
similar to that for BATSE, $P_{\rm lim}\sim 1\,{\rm ph\, cm^{-2}\,
s^{-1}}$, based on the observation that the detection rate of GRBs by
SWIFT ($\sim100\peryr$) is similar to that of BATSE (taking into
account the different field of views and the fact that BATSE was
triggering only 1/3 of the time), and that the fraction of SHBs is
similar for both BATSE and SWIFT. Fig. (\ref{cdf}) shows a comparison
between the observed and the expected integrated $z$-DF of SHBs for
the different models listed in table (~\ref{t:fit}). Our model for
dynamical DNS mergers fits the observed SHBs much better than the
model for primordial DNSs (see the results of Kolmogorov-Smirnoff [KS]
tests in table [\ref{t:fit}]).

We note, however, that ruling out a primordial binary $z$-DF may be
premature based on current data. For SWIFT, only 1/3 of detected SHBs
have secure $z$ determinations. If there is a bias against obtaining a
secure redshift for higher $z$ bursts, the observed distribution would
be shifted, compared to the expected distribution, to low
$z$. Assuming, for example, that obtaining a secure $z$ requires a
higher $P_{\rm lim}$ compared to that required for detection, the
lower rate of detection of SHBs with secure $z$ is accounted for by
choosing $P_{\rm lim}=2.5\,{\rm ph\,cm^{-2}\, s^{-1}}$ (which reduces
the detection rate by a factor of 3 compared to that obtained for
$P_{\rm lim}=1\,{\rm ph\,cm^{-2}\, s^{-1}}$). Indeed, inspection of
the 15 to 150~keV peak fluxes of SWIFT SHBs shows that the average
peak flux of SHBs with $z$ is higher than that of SHBs without $z$
($\langle f_z\rangle=5.7\pm 5.7\,{\rm ph\, cm^{-2}\, s^{-1}}$ for GRBs
050509, 050724 and 051221 compared to $\langle f_{{\rm no}\,
z}\rangle=1.2\pm 0.5\,{\rm ph\, cm^{-2}\,s^{-1}}$ for GRBs 050911,
051105, 051114, 051210 and 051227; Barthelmy et al. \cite{Bar05};
Gehrels et al. \cite{Gea05}; Bloom et al. \cite{Bea05}; Berger et
al. \cite{BPea05}; Fox et al. \cite{Fea05}; Villasenor et
al. \cite{Vil05}).

Choosing $P_{\rm lim}=2.5\,{\rm ph\,cm^{-2}\, s^{-1}}$, the expected
$z$-DF of primordial binary mergers is marginally compatible with
observations (see table~\ref{t:fit} and fig.~~\ref{cdf}).  The time
delay distribution $(dp/d\tau)_{\rm dyn}$ expected for dynamical DNSs
yields a $z$-DF compatible with observations for both choices of
$P_{\rm lim}$.

\section{Summary and Discussion}\label{s:discussion}

We have shown (fig~\ref{tdelay}) that the $z$-DF of SHBs expected from
{\it dynamical} formation and subsequent merger of DNSs is markedly
different from that expected from primordial DNS mergers (assuming
$(dp/d\tau)_{\rm prim}\propto 1/\tau$). The large time for core
collapse shifts the DF of dynamically formed DNS mergers to low
$z$. The observed $z$-DF of SHBs strongly favors that expected for
mergers of dynamically formed DNSs, as compared to that expected for
primordial DNS mergers (fig~\ref{cdf} and table~1; see also GP05 and
Nakar et al. \cite{NGF06}).

However, current data do not allow to rule out a $z$-DF consistent
with that expected for primordial DNS mergers, since redshifts were
obtained only for a minority of the detected SHBs. This may be due to
a bias against obtaining redshift information for high $z$ (faint)
SHBs (fig~\ref{cdf}, table~1 and discussion at the end of
\S3.3). Future observations should allow to better constrain the
$z$-DF of SHBs, and thus to differentiate between models. For example,
detection of only a few high redshift ($z>2$) SHBs would severely
constrain the contribution of dynamically formed DNSs.

If the formation rate of primordial and dynamically formed DNSs are
comparable (GPZM06), there will be an anti-correlation between $z$ and
off-set of the SHB from the center of its host galaxy, because DNSs
formed in GCs will be closer in $z$, but farther away from their host
center, since GCs reside in the halos of galaxies. The large time
delay also implies that more DNS mergers will be observed by LIGO and
VIRGO (Nakar et al. \cite{NGF06})

An alternative method for constraining the progenitors of SHBs is to
consider their demography (Gal-Yam et al \cite{Gal06}; Zheng \&
Ramirez-Ruiz \cite{Zhe06}). Zheng \& Ramirez-Ruiz (\cite{Zhe06})
showed that in order to account for the preponderance of SHBs in
elliptical galaxies, the delay times should be large, following
$dp/d\tau\propto \tau^\delta$ with $\delta\approx1.5$. This method,
which does not rely on the observed $z$-DF, also indicates delay times
longer than expected for primordial binaries. We note that the
specific frequency of GCs (number of GCs per unit luminosity in the
V-band) is larger for elliptical galaxies than for spirals
(e.g. Harris \cite{Har91}) by an order of magnitude.

\acknowledgements This work was supported by an AEC grant (EW), The
Netherlands Academy of Arts and Sciences (KNAW), The Netherlands
Organization for Scientific Research (NWO), and the The Netherlands
School for Astronomy (NOVA) (SPZ). DG thanks S. Covino, D. Malesani
and C. Guidorzi for useful information, and the Weizmann Inst. for its
hospitality. The research of DG is partially supported by MIUR-Cofin
grant.

\end{document}